\documentclass[amsmath,aps,showpacs,a4paper,10pt]{revtex4}

 \usepackage{epsf}
 \usepackage{graphicx}    

\begin{document}

 \newcommand{\be}[1]{\begin{equation}\label{#1}}
 \newcommand{\ee}{\end{equation}}
 \newcommand{\bea}{\begin{eqnarray}}
 \newcommand{\eea}{\end{eqnarray}}
 \def\disp{\displaystyle}

 \def\gsim{ \lower .75ex \hbox{$\sim$} \llap{\raise .27ex \hbox{$>$}} }
 \def\lsim{ \lower .75ex \hbox{$\sim$} \llap{\raise .27ex \hbox{$<$}} }

 \begin{titlepage}

 \begin{flushright}
 arXiv:1306.1364
 \end{flushright}

 \title{\Large \bf Indistinguishability of Warm Dark Matter,
 Modified Gravity,~and Coupled Cold Dark Matter}

 \author{Hao~Wei\,}
 \email[\,email address:\ ]{haowei@bit.edu.cn}
 \affiliation{School of Physics, Beijing Institute
 of Technology, Beijing 100081, China}

 \author{Jing~Liu}
 \affiliation{School of Physics, Beijing Institute
 of Technology, Beijing 100081, China}

 \author{Zu-Cheng~Chen\,}
 \affiliation{School of Physics, Beijing Institute
 of Technology, Beijing 100081, China}

 \author{Xiao-Peng~Yan\,}
 \affiliation{School of Physics, Beijing Institute
 of Technology, Beijing 100081, China}

 \begin{abstract}\vspace{1cm}
 \centerline{\bf ABSTRACT}\vspace{2mm}
 The current accelerated expansion of our universe could be due
 to an unknown energy component with negative pressure (dark
 energy) or a modification to general relativity (modified
 gravity). On the other hand, recently warm dark matter (WDM)
 remarkably rose as an alternative of cold dark matter (CDM).
 Obviously, it is of interest to distinguish these different
 types of models. In fact, many attempts have been made in
 the literature. However, in the present work, we show that
 WDM, modified gravity and coupled CDM form a trinity,
 namely, they are indistinguishable by using the cosmological
 observations of both cosmic expansion history and growth
 history. Therefore, to break this degeneracy, the other
 complementary probes beyond the ones of cosmic expansion
 history and growth history are required.
 \end{abstract}

 \pacs{98.80.-k, 95.36.+x, 95.35.+d, 04.50.-h}

 \maketitle

 \end{titlepage}

 \renewcommand{\baselinestretch}{1.1}


\section{Introduction}\label{sec1}

As is well known, the current accelerated expansion of
 our universe~\cite{r1} could be due to an unknown energy
 component with negative pressure (dark energy)~\cite{r1} or
 a modification to general relativity (modified
 gravity)~\cite{r1,r2}. So, it is of interest to distinguish
 these two types of models. However, most cosmological
 observations merely probe the expansion history of our
 universe. On the other hand, it is easy to build models
 that share the same cosmic expansion history by means of
 reconstruction. Thus, in order to distinguish various models,
 some independent and complementary probes are required.
 In fact, it was proposed that the measurement of growth
 function $\delta(z)$ might be capable (see
 e.g.~\cite{r3,r4,r5,r6,r30}). If the dark energy model and the
 modified gravity model share the same cosmic expansion
 history, their growth histories might be different, and hence
 they might be distinguished from each other.

However, up to now, in most works on this issue it is commonly
 assumed that there is no interaction between dark matter and
 dark energy. In fact, since the nature of both dark
 energy and dark matter are still unknown, there is no physical
 argument to exclude the possible interaction between them. On
 the contrary, some observational evidences of this possible
 interaction have been found. For instance,
 Bertolami {\it et al.}~\cite{r7} showed that the Abell Cluster
 A586 exhibits evidence of the interaction between dark energy
 and dark matter, and they argue that this interaction might
 imply a violation of the equivalence principle. On the other
 hand, Abdalla {\it et al.}~\cite{r8} found the signature of
 interaction between dark energy and dark matter by using
 optical, X-ray and weak lensing data from 33 relaxed galaxy
 clusters. Recently, Salvatelli {\it et al.}~\cite{r29} claimed
 that the Planck 2013 data favor a non-zero interaction between
 dark energy and dark matter. If the possible interaction
 between cold dark matter (CDM) and dark energy is allowed, it
 has been shown in e.g.~\cite{r9,r10} that the coupled CDM
 model can share both the same cosmic expansion history and
 growth history with the modified gravity model, and hence
 they cannot be distinguished.

On the other hand, although the well-known $\Lambda$CDM model
 is very successful in many aspects, it has been seriously
 challenged recently. According to the brief reviews in
 e.g.~\cite{r11}, these serious challenges include, for
 instance, (1)~$\Lambda$CDM predicts significantly smaller
 amplitude and scale of large-scale velocity flows than
 observations; (2) $\Lambda$CDM predicts fainter Type Ia
 supernova (SNIa) at high redshift $z$; (3) $\Lambda$CDM
 predicts more dwarf or irregular galaxies in voids than
 observed; (4) $\Lambda$CDM predicts shallow low concentration
 and density profiles of cluster haloes in contrast to
 observations; (5) $\Lambda$CDM predicts galaxy halo mass
 profiles with cuspy cores and low outer density while
 observations indicate a central core of constant density and
 a flattish high dark mass density outer profile;
 (6) $\Lambda$CDM predicts a smaller fraction of disk galaxies
 due to recent mergers expected to disrupt cold rotationally
 supported disks. Even when one replaces the cosmological
 constant $\Lambda$ with other (dynamical) dark energy
 candidates, these challenges still cannot be successfully
 addressed. In particular, the main source of the challenges
 on the small/galactic scale might be CDM. We refer to
 e.g.~\cite{r11} for details.

Recently, warm dark matter (WDM) remarkably rose as an
 alternative of CDM. The leading WDM candidates are the keV
 scale sterile neutrinos. In fact, the keV scale WDM is an
 intermediate case between the eV scale hot dark matter (HDM)
 and the GeV scale CDM. Unlike CDM which is challenged on the
 small/galactic scale (as mentioned above), it is claimed that
 WDM can successfully reproduce the astronomical observations
 over all the scales (from small/galactic to large/cosmological
 scales)~\cite{r12}. The key is the connection between the
 mass of dark matter (DM) particles and the free-streaming
 length $\ell_{\rm fs}$ (structure smaller than
 $\ell_{\rm fs}$ will be erased). The eV scale HDM is too
 light, and hence all structures below the Mpc scale will be
 erased; the GeV scale CDM is too heavy, and hence the
 structures below the kpc scale cannot be erased (therefore,
 CDM is challenged on the small/galactic scale).
 In between, the keV scale WDM works well~\cite{r12}. We refer
 to e.g.~\cite{r12} for a comprehensive review.

WDM has a fairly small but non-zero equation-of-state parameter
 (EoS), while the EoS of CDM is zero. In the literature
 (e.g.~\cite{r13,r14,r15,r16}), many attempts have been made to
 constrain the EoS of WDM $w_m$, and it was found that $w_m$ is
 about ${\cal O}(10^{-3})\sim{\cal O}(10^{-2})$ by using the
 current cosmological data. Of course, $w_m$ is not constant
 in general. Let us consider the energy conservation equation
 of WDM, namely
 \be{eq1}
 \dot{\rho}_m+3H\rho_m\left(1+w_m\right)=0\,,
 \ee
 where a dot denotes a derivative with respect to cosmic time
 $t$; $H\equiv\dot{a}/a$ is the Hubble parameter;
 $a=(1+z)^{-1}$ is the scale factor (we have set $a_0=1$; the
 subscript ``0'' indicates the present value of corresponding
 quantity; $z$ is the redshift); $\rho_m$ is the energy
 density of WDM (we assume that the baryon component is
 negligible). One can naively rewrite Eq.~(\ref{eq1}) as
 \be{eq2}
 \dot{\rho}_m+3H\rho_m=-\Gamma^{\rm eff}\,,~~~~~~~{\rm and}
 ~~~~~~~\Gamma^{\rm eff}=3H\rho_m w_m\not=0\,.
 \ee
 Obviously, Eq.~(\ref{eq2}) could be regarded as the energy
 conservation equation of coupled CDM, while the
 term $\Gamma^{\rm eff}$ could be regarded as the interaction
 between CDM and dark energy. Correspondingly, the original
 energy conservation equation of dark energy
 $\dot{\rho}_{de}+3H\rho_{de}\left(1+w_{de}\right)=0$ can
 be rewritten as
 \be{eq3}
 \dot{\rho}_{de}+3H\rho_{de}\left(1+w_{de}^{\rm eff}\right)
 =\Gamma^{\rm eff}\,,~~~~~~~{\rm and}~~~~~~~
 w_{de}^{\rm eff}=w_{de}+\Gamma^{\rm eff}/(3H\rho_{de})\,.
 \ee
 Obviously, this procedure can be reversed. So, we can naively
 see that WDM could be equivalent to coupled CDM in this sense.
 Of course, it should be checked that WDM and coupled CDM can
 share both the same cosmic expansion history and growth
 history before we can say they are really indistinguishable.
 This will be part of our goal of this work. As mentioned
 above, in e.g.~\cite{r9,r10} it was shown that modified
 gravity and CDM coupled with dark energy are indistinguishable
 since they can share both the same cosmic expansion history
 and growth history. Thus, it is very natural to speculate that
 modified gravity and WDM are also indistinguishable. In fact,
 the main goal of this work is to check the idea mentioned
 here. If these three types of cosmological models are really
 indistinguishable, some complementary probes beyond the ones
 of cosmic expansion history and growth history are required.

Before we go further, it is important to clarify a pitfall
 in the above naive discussions. In fact, the route from
 Eq.~(\ref{eq1}) to Eq.~(\ref{eq2}) is not unique. One can
 rather rewrite Eq.~(\ref{eq1}) as
 \be{eq4}
 \dot{\rho}_m^{\rm eff}+3H\rho_m^{\rm eff}=-\Gamma^{\rm eff}\,,
 ~~~~~~~\rho_m^{\rm eff}=\rho_m+X\,,~~~~~~~~
 \Gamma^{\rm eff}=3H\rho_m w_m-\dot{X}-3HX\,,
 \ee
 where $X$ can be any quantity. Obviously,
 $\rho_m^{\rm eff}\not=\rho_m$ and
 $\Gamma^{\rm eff}\not=3H\rho_m w_m$ in general. So, the energy
 density of coupled CDM $\hat{\rho}_m$ is not necessarily
 equal to the one of WDM $\rho_m$, and hence the fractional
 energy density $\hat{\Omega}_m\not=\Omega_m$ in general.

In Sec.~\ref{sec2}, we firstly consider modified gravity and
 WDM. We propose a general approach to construct a WDM model
 that shares both the same expansion history and growth
 history with modified gravity. Then, an explicit example will
 be shown. In Sec.~\ref{sec3}, we turn to WDM and coupled CDM.
 Also, we propose a general approach to construct a coupled
 CDM model that shares both the same expansion history and
 growth history with the WDM model. Of course, we will show an
 explicit example, too. Finally, some brief concluding remarks
 are given in Sec.~\ref{sec4}.


\section{Modified gravity and WDM}\label{sec2}


\subsection{General formalism}\label{sec2a}

Throughout this work, we consider a flat Friedmann-Robertson-Walker
 (FRW) universe. We firstly consider modified gravity and WDM.
 In this section, for the WDM model, we assume that the universe
 contains only WDM and dark energy (note that in general
 relativity, it is required that dark energy coexists with
 WDM to accelerate the cosmic expansion). The Friedmann
 equation reads
 \be{eq5}
 H^2=\frac{8\pi G}{3}\left(\rho_m+\rho_{de}\right)\,,
 \ee
 where $\rho_m$ and $\rho_{de}$ are the energy densities of WDM
 and dark energy, respectively. In the WDM model under
 consideration, we assume that there is no interaction between
 WDM and dark energy, and hence their energy conservation
 equations are given by
 \bea
 &&\dot{\rho}_m+3H\rho_m\left(1+w_m\right)=0\,,\label{eq6}\\
 &&\dot{\rho}_{de}+3H\rho_{de}\left(1+w_{de}\right)=0\,,\label{eq7}
 \eea
 where $w_m$ and $w_{de}$ are the EoS of WDM and dark energy,
 respectively. From Eqs.~(\ref{eq6}) and (\ref{eq7}), we have
 \bea
 &&\Omega_m^\prime=-\Omega_m\left[\,3\left(1+w_m\right)+
 2\frac{H^\prime}{H}\,\right]\,,\label{eq8}\\
 &&\Omega_{de}^\prime=-\Omega_{de}\left[\,3\left(1+w_{de}\right)+
 2\frac{H^\prime}{H}\,\right]\,,\label{eq9}
 \eea
 where a prime denotes a derivative with respect to $\ln a$,
 and $\Omega_i\equiv 8\pi G\rho_i/(3H^2)$ are the fractional
 energy densities for WDM and dark energy. On the side of
 growth history, in general relativity, the perturbation
 equation for WDM in the sub-horizon regime
 is given by (see e.g.~\cite{r17,r18,r19})
 \be{eq10}
 \ddot{\delta}+\left[2-3\left(2w_m-c_s^2\right)\right]H\dot{\delta}
 =4\pi G\rho_m\delta\left(1-6c_s^2+8w_m-3w_m^2\right)\,,
 \ee
 where $\delta\equiv\delta\rho_m/\rho_m$ is the linear matter
 density contrast, and $c_s^2\equiv\dot{p}_m/\dot{\rho}_m$ is
 the sound speed squared of WDM. Using $p_m=w_m\rho_m$ and
 Eq.~(\ref{eq6}), we find that
 \be{eq11}
 c_s^2=w_m-\frac{w_m^\prime}{3(1+w_m)}\,.
 \ee
 Obviously, if $w_m=const.$, we have $c_s^2=w_m=const.$.
 Therefore, for CDM (namely $w_m=0$), it is easy to see that
 Eq.~(\ref{eq10}) reduces to the well-known form in general
 relativity~\cite{r3,r4,r5,r6,r9,r10,r30}
 \be{eq12}
 \ddot{\delta}+2H\dot{\delta}=4\pi G\rho_m\delta\,.
 \ee
 For convenience, we recast Eq.~(\ref{eq10}) as
 \be{eq13}
 \delta^{\prime\prime}+\left[\,2-3\left(2w_m-c_s^2\right)+
 \frac{H^\prime}{H}\,\right]\delta^\prime=\frac{3}{2}\Omega_m
 \delta\left(1-6c_s^2+8w_m-3w_m^2\right)\,.
 \ee

On the other hand, in modified gravity, the perturbation
 equation reads~\cite{r4,r20,r21}
 \be{eq14}
 \ddot{\tilde{\delta}}+2\tilde{H}\dot{\tilde{\delta}}
 =4\pi G_{\rm eff}\tilde{\rho}_m\tilde{\delta}\,,
 \ee
 where the quantities in modified gravity are labeled by a
 tilde ``$\sim$''; $G_{\rm eff}$ is the effective local
 gravitational ``constant'' measured by Cavendish-type
 experiment, which is time-dependent. Note that in the
 modified gravity model under consideration, the role of
 dark matter is played by CDM (its EoS is zero). In general,
 $G_{\rm eff}$ can be written as~\cite{r4,r20,r21}
 \be{eq15}
 G_{\rm eff}=G\left(1+\frac{1}{3\beta}\right)\,,
 \ee
 where $\beta$ is a conventional quantity introduced just for
 convenience (which is equivalent to $G_{\rm eff}$ in fact).
 $\beta$ will be determined once we specify the modified
 gravity model (see below). Eq.~(\ref{eq14}) can be recast as
 \be{eq16}
 \tilde{\delta}^{\prime\prime}+\left(2
 +\frac{\tilde{H}^\prime}{\tilde{H}}\right)
 \tilde{\delta}^\prime=\frac{3}{2}\left(1+
 \frac{1}{3\beta}\right)\tilde{\Omega}_m\tilde{\delta}\,.
 \ee

To be indistinguishable, we require that the WDM model shares
 both the same expansion history and growth history with the
 modified gravity model, namely, we identify
 \be{eq17}
 H=\tilde{H}~~~{\rm and}~~~\delta=\tilde{\delta}\,.
 \ee
 Comparing Eq.~(\ref{eq13}) with Eq.~(\ref{eq16}), it is
 easy to find that
 \be{eq18}
 2\left(2w_m-c_s^2\right)\tilde{\delta}^\prime=\tilde{\delta}
 \left[\tilde{\Omega}_m\left(1+\frac{1}{3\beta}\right)-
 \Omega_m\left(1-6c_s^2+8w_m-3w_m^2\right)\right]\,.
 \ee
 Note that $\Omega_m\not=\tilde{\Omega}_m$ in general.

Let us describe the prescription to construct the WDM model
 that shares both the same expansion history and growth
 history with a given modified gravity model. For a given
 modified gravity model, all its $\tilde{\Omega}_m$, $\beta$,
 and $\tilde{H}$ are known. Then, we can solve the
 differential equation~(\ref{eq16}) and obtain $\tilde{\delta}$
 as a function of $\ln a$. Note that $H=\tilde{H}$ and
 $\delta=\tilde{\delta}$. Substituting Eq.~(\ref{eq11}) into
 Eq.~(\ref{eq18}), we find that Eqs.~(\ref{eq18})
 and~(\ref{eq8}) become two coupled first-order differential
 equations for $w_m$ and $\Omega_m$ with respect to $\ln a$.
 Obviously, we can numerically solve them and get $w_m$ and
 $\Omega_m$ as functions of $\ln a$. Then,
 $\Omega_{de}=1-\Omega_m$ and $c_s^2$ in Eq.~(\ref{eq11}) are
 on hand. From Eq.~(\ref{eq9}), we obtain the EoS of dark
 energy, namely
 \be{eq19}
 w_{de}=-1-\frac{1}{3}\left(\frac{\Omega_{de}^\prime}{\Omega_{de}}+
 2\frac{H^\prime}{H}\right)\,,
 \ee
 in which we note that $H=\tilde{H}$ from Eq.~(\ref{eq17}).
 Of course, if one needs $\rho_i=3H^2\Omega_i/(8\pi G)$ for
 WDM and dark energy, noting $H=\tilde{H}$, they are also
 ready. So, all the physical quantities of the corresponding
 WDM model are known. The resulting WDM model is really
 indistinguishable from the given modified gravity model
 by using the observations of both expansion history and
 growth history.


\subsection{Explicit example}\label{sec2b}

\begin{center}
 \begin{figure}[htbp]
 \centering
 \includegraphics[width=0.85\textwidth]{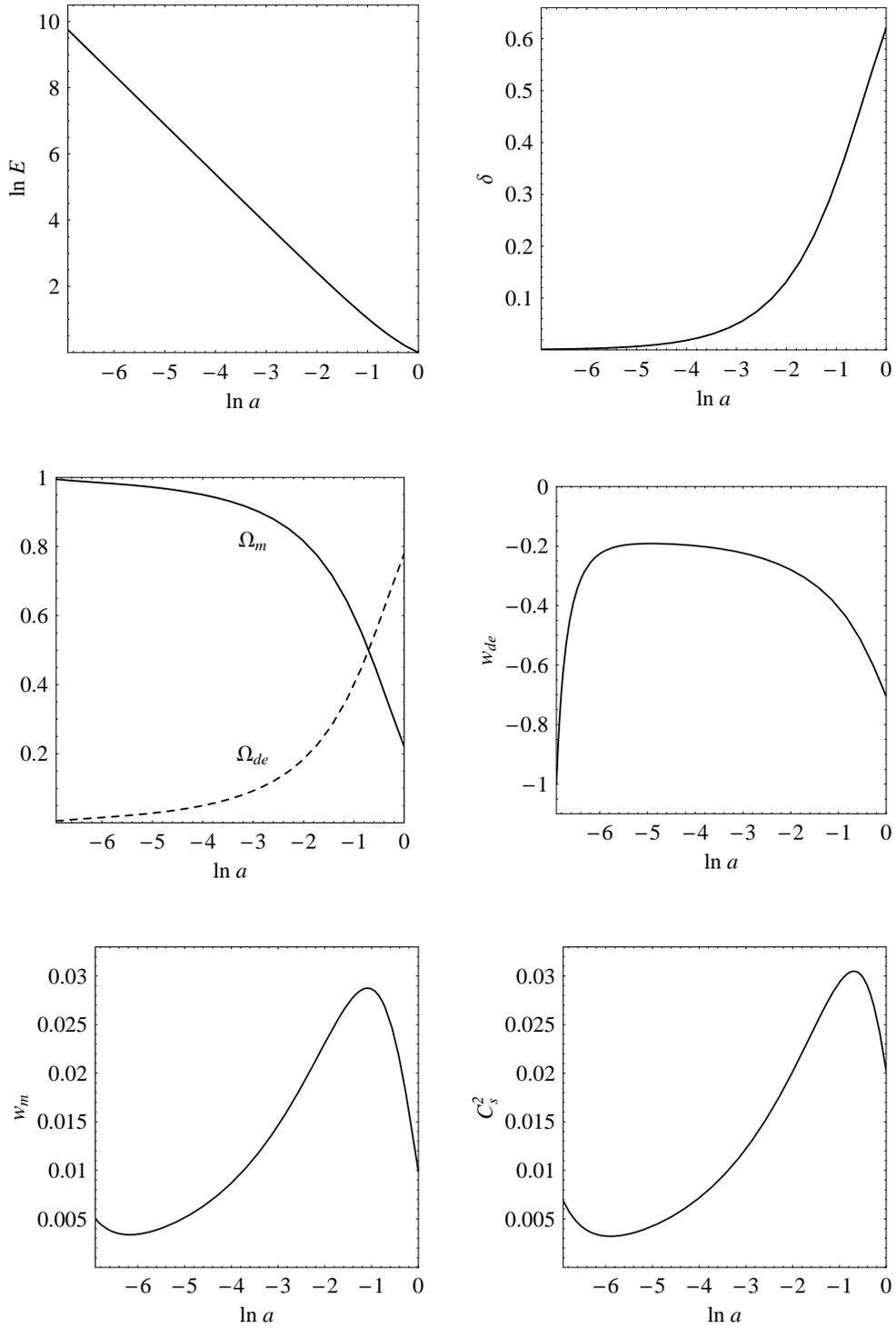}
 \caption{\label{fig1}
 $\ln E=\ln\tilde{E}$, $\delta=\tilde{\delta}$, $\Omega_m$,
 $\Omega_{de}$, $w_{de}$, $w_m$ and $c_s^2$ as functions of
 $\ln a$. See the text for details.}
 \end{figure}
 \end{center}

Here, we would like to give an explicit example following the
 prescription proposed in Sec.~\ref{sec2a}. As an example, we
 consider the simplest modified gravity model, namely, the
 flat Dvali-Gabadadze-Porrati~(DGP) braneworld model~\cite{r22}
 (see also e.g.~\cite{r4,r20,r21,r23}). Note that here we only
 consider the self-accelerating branch of DGP model, for which
 the reduced Hubble parameter is given by~\cite{r4,r22,r23}
 \be{eq20}
 \tilde{E}\equiv\frac{\tilde{H}}{\tilde{H}_0}=
 \sqrt{\tilde{\Omega}_{m0}(1+z)^3+\tilde{\Omega}_{rc}}+
 \sqrt{\tilde{\Omega}_{rc}}=\sqrt{\tilde{\Omega}_{m0}\,
 e^{-3\ln a}+\tilde{\Omega}_{rc}}+\sqrt{\tilde{\Omega}_{rc}}\,,
 \ee
 where $\tilde{\Omega}_{rc}$ is a constant.
 Requiring $\tilde{E}(z=0)=1$ by definition, it is easy to see that
 \be{eq21}
 \tilde{\Omega}_{m0}=1-2\sqrt{\tilde{\Omega}_{rc}}\,.
 \ee
 Thus, the DGP model has only an independent model parameter,
 namely $\tilde{\Omega}_{rc}\,$. Note that
 $0\leq\tilde{\Omega}_{rc}\leq 1/4$ is required
 by $0\leq \tilde{\Omega}_{m0}\leq 1$. The fractional energy density
 of dark matter in the DGP model is given by~\cite{r4,r22,r23}
 \be{eq22}
 \tilde{\Omega}_m=\frac{\tilde{\Omega}_{m0}(1+z)^3}{\tilde{E}^2(z)}
 =\frac{\tilde{\Omega}_{m0}\,e^{-3\ln a}}{\tilde{E}^2(\ln a)}.
 \ee
 On the other hand, the $\beta$ in Eq.~(\ref{eq15}) for the
 flat DGP model is given by~\cite{r4,r20,r21}
 \be{eq23}
 \beta=-\frac{1+\tilde{\Omega}_m^2}{1-\tilde{\Omega}_m^2}.
 \ee
 For demonstration, we choose the single independent model
 parameter as $\tilde{\Omega}_{rc}=0.125$, which is well
 consistent with the current observational data
 (see e.g.~\cite{r24,r25}). Substituting this
 given $\tilde{\Omega}_{rc}$ into Eq.~(\ref{eq21}) and
 then Eqs.~(\ref{eq20}), (\ref{eq22}) and (\ref{eq23}), the
 corresponding $\tilde{E}$, $\tilde{\Omega}_m$, and $\beta$
 as functions of $\ln a$ are known. Noting that
 $H^\prime/H=\tilde{H}^\prime/\tilde{H}=\tilde{E}^\prime/\tilde{E}$,
 and following the prescription proposed in Sec.~\ref{sec2a},
 we can easily construct the desired WDM model which shares
 both the same expansion history and growth history with this
 given DGP model. Firstly, we obtain $\delta=\tilde{\delta}$
 by solving Eq.~(\ref{eq16}). As is well known,
 $\tilde{\delta}^\prime=\tilde{\delta}=a$ at $z\gg 1$
 (see e.g.~\cite{r3,r4,r5,r6}). Therefore, we use the initial
 condition $\tilde{\delta}^\prime=\tilde{\delta}=a_{ini}$ at
 $z_{ini}=1000$ for the differential equation~(\ref{eq16}).
 The resulting $\delta=\tilde{\delta}$ and $\ln E=\ln\tilde{E}$
 as functions of $\ln a$ are shown in Fig.~\ref{fig1}.
 Secondly, substituting Eq.~(\ref{eq11}) into Eq.~(\ref{eq18}),
 noting
 $H^\prime/H=\tilde{H}^\prime/\tilde{H}=\tilde{E}^\prime/\tilde{E}$,
 Eqs.~(\ref{eq18}) and (\ref{eq8}) become two
 coupled first-order differential equations for $w_m$ and
 $\Omega_m$ with respect to $\ln a$. We can numerically solve
 them with the demonstrative initial conditions
 $\Omega_m(z=z_{ini})=0.995$, $w_m(z=z_{ini})=0.005$ (which is
 well consistent with the observational data~\cite{r16}),
 and then obtain $w_m$, $\Omega_m$ as functions of $\ln a$.
 They are also shown in Fig.~\ref{fig1}. Finally,
 $\Omega_{de}=1-\Omega_m$, as well as $c_s^2$ in Eq.~(\ref{eq11}) and
 $w_{de}$ in Eq.~(\ref{eq19}) are available, and they can also
 be found in Fig.~\ref{fig1}. So far, we have successfully
 constructed the WDM model that shares both the same expansion
 history and growth history with a given modified gravity
 model, namely, the flat DGP model. These two models are
 indistinguishable in this sense. Therefore, to distinguish the
 modified gravity model and WDM model, it is required to seek
 some complementary probes beyond the ones of cosmic expansion
 history and growth history (for instance, the observations on
 the small/galactic scale).


\vspace{-7mm} 


\section{WDM and coupled CDM}\label{sec3}


\subsection{General formalism}\label{sec3a}

In this section, we turn to WDM and coupled CDM. Since we have
 shown that WDM and modified gravity are indistinguishable in
 Sec.~{\ref{sec2}, and we have already shown that modified
 gravity and coupled CDM are indistinguishable in~\cite{r9}
 (see also e.g.~\cite{r10}), it is reasonable to expect that
 WDM and coupled CDM are also indistinguishable. Here, we try
 to construct a coupled CDM model that shares both the same
 expansion history and growth history with the WDM model.

Following~\cite{r9}, we consider the case of CDM coupled with
 quintessence (which is a main candidate of dark energy). It
 is well known that the pressure and energy density for the
 homogeneous quintessence are given by
 \be{eq24}
 \hat{p}_\phi=\frac{1}{2}\dot{\phi}^2-V(\phi)\,,~~~~~~~
 \hat{\rho}_\phi=\frac{1}{2}\dot{\phi}^2+V(\phi)\,,
 \ee
 where $V$ is the potential of the scalar field $\phi$, and the
 quantities in coupled CDM model are labeled by a hat ``$\wedge$''.
 The corresponding Friedmann equation reads
 \be{eq25}
 3\hat{H}^2=\kappa^2\left(\hat{\rho}_m+\hat{\rho}_\phi\right)\,,
 \ee
 where $\kappa^2\equiv 8\pi G$. We assume that CDM and quintessence
 interact through~\cite{r26,r27,r28}
 \bea
 &&\dot{\hat{\rho}}_m+3\hat{H}\hat{\rho}_m=-\kappa
 Q\hat{\rho}_m\dot{\phi}\,,\label{eq26}\\
 &&\dot{\hat{\rho}}_\phi+3\hat{H}\left(\hat{\rho}_\phi
 +\hat{p}_\phi\right)=\kappa Q\hat{\rho}_m\dot{\phi}\,,\label{eq27}
 \eea
 which preserve the total energy conservation equation
 $\dot{\hat{\rho}}_{tot}+3\hat{H}\left(\hat{\rho}_{tot}
 +\hat{p}_{tot}\right)=0$. The dimensionless
 coupling coefficient $Q=Q(\phi)$ is an arbitrary function
 of $\phi$. In fact, Eq.~(\ref{eq27}) is equivalent to
 \be{eq28}
 \ddot{\phi}+3\hat{H}\dot{\phi}
 +\frac{dV}{d\phi}=\kappa Q\hat{\rho}_m\,.
 \ee
 Using Eqs.~(\ref{eq25}), (\ref{eq26}) and (\ref{eq27}), one
 can obtain the Raychaudhuri equation, namely
 \be{eq29}
 \dot{\hat{H}}=-\frac{\kappa^2}{2}\left(\hat{\rho}_m+\hat{\rho}_\phi
 +\hat{p}_\phi\right)=-\frac{\kappa^2}{2}\left(\hat{\rho}_m+
 \dot{\phi}^2\right)\,.
 \ee
 It is worth noting that $\hat{\rho}_m$ does not scale as
 $a^{-3}$ due to the non-vanishing interaction. On the side
 of growth history, the perturbation equation in
 the sub-horizon regime is given by~\cite{r27}
 \be{eq30}
 \hat{\delta}^{\prime\prime}+\left(2+\frac{\hat{H}^\prime}{\hat{H}}
 -\kappa Q\phi^\prime\right)\hat{\delta}^\prime
 =\frac{3}{2}\left(1+2Q^2\right)\hat{\Omega}_m\hat{\delta}\,.
 \ee
 Obviously, if $Q=0$, Eq.~(\ref{eq30}) reduces to
 the well-known form (n.b.~Eq.~(\ref{eq12})). In
 fact, Eq.~(\ref{eq30}) from~\cite{r27} is valid for any
 $Q=Q(\phi)$ and generalizes the one of~\cite{r28} which is
 only valid for constant $Q$.


 \begin{center}
 \begin{figure}[htbp]
 \centering
 \includegraphics[width=0.85\textwidth]{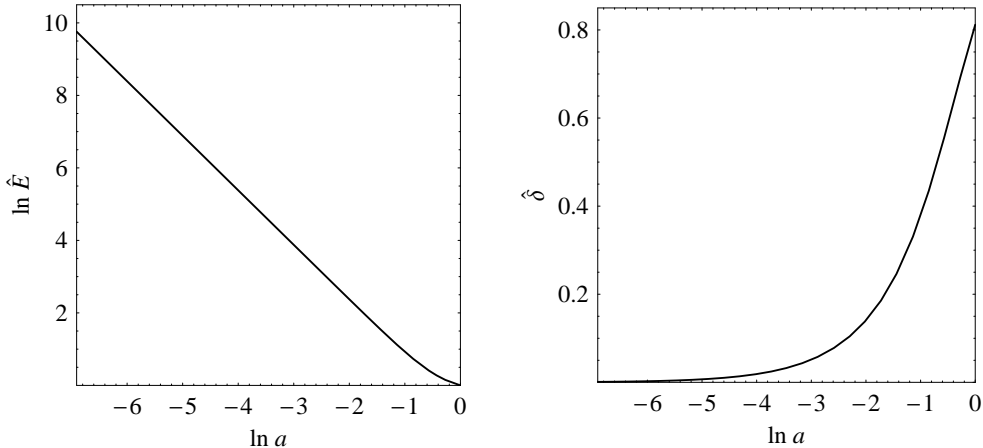}
 \caption{\label{fig2}
 $\ln\hat{E}=\ln E$ and $\hat{\delta}=\delta$ as
 functions of $\ln a$. See the text for details.}
 \end{figure}
 \end{center}


\vspace{-10.5mm} 

On the other hand, the perturbation equation in the WDM
 model has already been given in Eq.~(\ref{eq13}). To be
 indistinguishable, we require that the coupled CDM model
 shares both the same expansion history and growth
 history with the WDM model, namely, we identify
 \be{eq31}
 H=\hat{H}~~~{\rm and}~~~\delta=\hat{\delta}\,.
 \ee
 Comparing Eq.~(\ref{eq13}) with Eq.~(\ref{eq30}), it is
 easy to find that
 \be{eq32}
 \left[\,3\left(2w_m-c_s^2\right)-\kappa Q\phi^\prime\,\right]
 \delta^\prime=\frac{3}{2}\delta\left[\hat{\Omega}_m\left(
 1+2Q^2\right)-\Omega_m\left(1-6c_s^2+8w_m-3w_m^2
 \right)\right]\,.
 \ee
 Note that $\hat{\Omega}_m\not=\Omega_m$ in general, as
 mentioned in Sec.~\ref{sec1}. From Eq.~(\ref{eq29}), we have
 \be{eq33}
 \left(\kappa\phi^\prime\right)^2=-3\hat{\Omega}_m
 -2\frac{\hat{H}^\prime}{\hat{H}}\,.
 \ee
 From Eq.~(\ref{eq26}), it is easy to obtain
 \be{eq34}
 \hat{\Omega}_m^\prime=-\left(3+2\frac{\hat{H}^\prime}{\hat{H}}
 +\kappa Q\phi^\prime\right)\hat{\Omega}_m\,.
 \ee
 It turns out that
 \be{eq35}
 \kappa Q\phi^\prime=-3-2\frac{\hat{H}^\prime}{\hat{H}}
 -\frac{\hat{\Omega}_m^\prime}{\hat{\Omega}_m}\,.
 \ee
 From Eqs.~(\ref{eq33}) and~(\ref{eq35}), we have
 \be{eq36}
 Q^2=\frac{\left(\kappa Q\phi^\prime\right)^2}
 {\left(\kappa\phi^\prime\right)^2}
 =\left(3+2\frac{\hat{H}^\prime}{\hat{H}}+
 \frac{\hat{\Omega}_m^\prime}{\hat{\Omega}_m}\right)^2
 \left(-3\hat{\Omega}_m-2\frac{\hat{H}^\prime}{\hat{H}}\right)^{-1}.
 \ee
 From Eqs.~(\ref{eq24}), (\ref{eq25}) and (\ref{eq33}), we
 obtain the dimensionless potential of quintessence, namely
 \be{eq37}
 U\equiv\frac{\kappa^2 V}{\hat{H}_0^2}=3\hat{E}^2
 \left(1-\frac{\hat{\Omega}_m}{2}+\frac{1}{3}
 \frac{\hat{H}^\prime}{\hat{H}}\right)\,,
 \ee
 where $\hat{E}\equiv\hat{H}/\hat{H}_0$. Obviously, $U$
 is equivalent to $V$ in fact. From Eqs.~(\ref{eq24}),
 (\ref{eq33}) and (\ref{eq37}), we find the
 EoS of quintessence, namely
 \be{eq38}
 w_\phi=\frac{p_\phi}{\rho_\phi}=
 \left(-1-\frac{2}{3}\frac{\hat{H}^\prime}{\hat{H}}\right)
 \left(1-\hat{\Omega}_m\right)^{-1}.
 \ee

Let us describe the prescription to construct the coupled
 CDM model which shares both the same expansion history and
 growth history with a given WDM model. For a given WDM model,
 all its $w_m$, $c_s^2$, $\Omega_m$, and $H$ are known. Then,
 we can solve the differential equation~(\ref{eq13}) and obtain
 $\delta$ as a function of $\ln a$. Note that $\hat{H}=H$ and
 $\hat{\delta}=\delta$. Substituting $\delta$,
 Eqs.~(\ref{eq35}) and (\ref{eq36}) into Eq.~(\ref{eq32}),
 noting $\hat{H}=H$, Eq.~(\ref{eq32}) becomes a first-order
 differential equation for $\hat{\Omega}_m$ with respect to
 $\ln a$. Obviously, we can numerically solve this differential
 equation and get $\hat{\Omega}_m$ as a function of $\ln a$.
 So, $\hat{\Omega}_\phi=1-\hat{\Omega}_m$ is available.
 Substituting $\hat{\Omega}_m$ into Eqs.~(\ref{eq33}),
 (\ref{eq35}), (\ref{eq37}) and (\ref{eq38}),
 noting $\hat{H}=H$, we find $\kappa\phi^\prime$,
 $\kappa Q\phi^\prime$, $U$ and $w_\phi$ as functions of
 $\ln a$. Then, $Q=(\kappa Q\phi^\prime)/(\kappa\phi^\prime)$
 is on hand. By integrating $\kappa\phi^\prime$, we get
 $\kappa\phi$ as a function of $\ln a$. Therefore, we can
 finally obtain $Q$ and $U$ as functions of $\kappa\phi$.
 So, all the physical quantities of the corresponding
 coupled CDM model are known. The resulting coupled CDM model
 is really indistinguishable from the given WDM model by using
 the observations of both expansion history and growth history.


 \begin{center}
 \begin{figure}[htbp]
 \centering
 \includegraphics[width=0.85\textwidth]{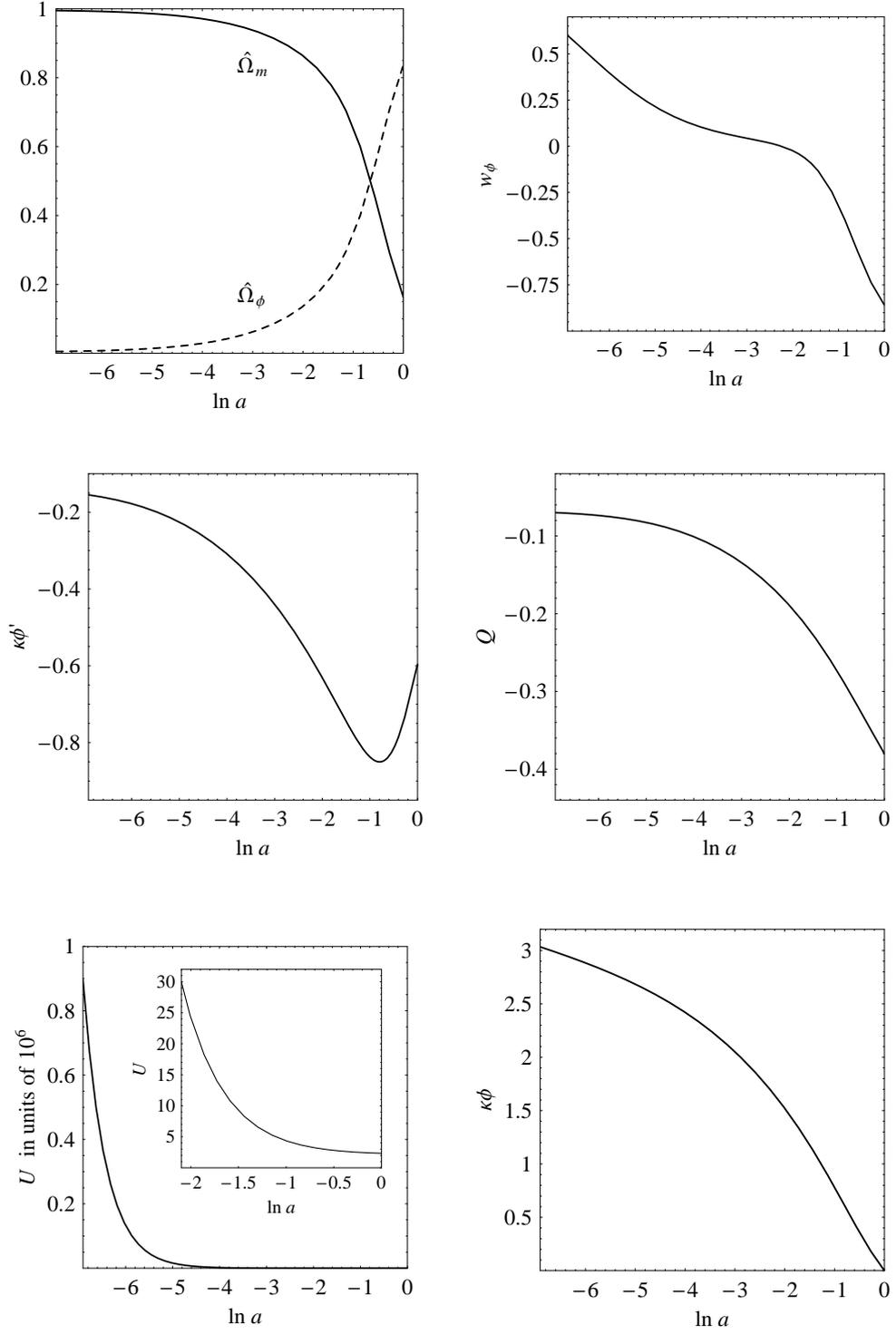}
 \caption{\label{fig3}
 $\hat{\Omega}_m$, $\hat{\Omega}_\phi$, $w_\phi$,
 $\kappa\phi^\prime$, $Q$, $U$ and $\kappa\phi$
 as functions of $\ln a$. See the text for details.}
 \end{figure}
 \end{center}


\vspace{-10.5mm} 


\subsection{Explicit example}\label{sec3b}

Here, we give an explicit example following the prescription
 proposed in Sec.~\ref{sec3a}. As an example, we consider the
 simplest WDM model, namely the so-called $\Lambda$WDM model,
 in which the role of dark energy is played by a cosmological
 constant $\Lambda$, while the EoS of WDM $w_m$ is a constant.
 In this case, from Eq.~(\ref{eq11}), we have
 $c_s^2=w_m=const.$. As is well known, the corresponding
 reduced Hubble parameter of the $\Lambda$WDM model reads (see
 e.g.~\cite{r16})
 \be{eq39}
 E\equiv\frac{H}{H_0}=\left[\Omega_{m0}(1+z)^{3(1+w_m)}
 +\left(1-\Omega_{m0}\right)\right]^{1/2}=
 \left[\Omega_{m0}\,e^{-3(1+w_m)\ln a}+\left(1-\Omega_{m0}
 \right)\right]^{1/2}.
 \ee
 There are two independent model parameters. For demonstration,
 we choose the model parameters of $\Lambda$WDM as
 $\Omega_{m0}=0.28$ and $w_m=0.003$, which is well
 consistent with the observational data (see e.g.~\cite{r16}).
 On the other hand, the fractional energy density of WDM is
 given by
 \be{eq40}
 \Omega_m\equiv\frac{8\pi G\rho_m}{3H^2}=
 \frac{\Omega_{m0}\,e^{-3(1+w_m)\ln a}}{E^2(\ln a)}\,.
 \ee
 Following the prescription proposed in Sec.~\ref{sec3a},
 we can easily construct the desired coupled CDM model that
 shares both the same expansion history and growth history
 with this given WDM model. Firstly, substituting
 Eqs.~(\ref{eq39}) and (\ref{eq40}) into Eq.~(\ref{eq13}),
 noting $H^\prime/H=E^\prime/E$, we can numerically solve
 this differential equation~(\ref{eq13}) and get $\delta$
 as a function of $\ln a$. As is well known,
 $\delta^\prime=\delta=a$ at $z\gg 1$ (see
 e.g.~\cite{r3,r4,r5,r6}). Therefore, we use the initial
 condition $\delta^\prime=\delta=a_{ini}$ at $z_{ini}=1000$
 for the differential equation~(\ref{eq13}). The resulting
 $\hat{\delta}=\delta$ and $\ln\hat{E}=\ln E$ as functions
 of $\ln a$ are shown in Fig.~\ref{fig2}.
 Secondly, substituting Eqs.~(\ref{eq35}), (\ref{eq36})
 and (\ref{eq40}) into Eq.~(\ref{eq32}), noting
 $\hat{H}^\prime/\hat{H}=H^\prime/H=E^\prime/E$, it is easy to
 see that Eq.~(\ref{eq32}) becomes a first-order differential
 equation for $\hat{\Omega}_m$ with respect to $\ln a$.
 Obviously, we can numerically solve this differential
 equation with the demonstrative initial condition
 $\hat{\Omega}_m(z=z_{ini})=0.995$, and get $\hat{\Omega}_m$
 as a function of $\ln a$. We show the resulting
 $\hat{\Omega}_m$ and $\hat{\Omega}_\phi=1-\hat{\Omega}_m$
 in Fig.~\ref{fig3}. Substituting $\hat{\Omega}_m$ into
 Eqs.~(\ref{eq33}), (\ref{eq35}), (\ref{eq37}) and
 (\ref{eq38}), noting
 $\hat{H}^\prime/\hat{H}=H^\prime/H=E^\prime/E$, we find
 $\kappa\phi^\prime$, $\kappa Q\phi^\prime$, $U$ and $w_\phi$
 as functions of $\ln a$. Then,
 $Q=(\kappa Q\phi^\prime)/(\kappa\phi^\prime)$ is on hand.
 By integrating $\kappa\phi^\prime$, we get $\kappa\phi$ as
 a function of $\ln a$. Note that for demonstration, we
 choose the negative branch for $\kappa\phi^\prime$, and
 choose $\phi_0=0$ when we get $\kappa\phi$. In
 Fig.~\ref{fig3}, we also show the resulting $w_\phi$,
 $\kappa\phi^\prime$, $Q$, $U$ and $\kappa\phi$ as functions
 of $\ln a$. Once we obtain $Q$, $U$ and $\kappa\phi$ as
 functions of $\ln a$, it is easy to find $Q$ and $U$ as
 functions of $\kappa\phi$. The results are shown
 in Fig.~\ref{fig4}. So far, we have successfully
 constructed the coupled CDM model that shares both the same
 expansion history and growth history with a given WDM model,
 namely, the $\Lambda$WDM model. These two models are
 indistinguishable in this sense. Therefore, to distinguish
 the coupled CDM model and WDM model, it is required to seek
 some complementary probes beyond the ones of cosmic expansion
 history and growth history (for instance, the observations on
 the small/galactic scale).


 \begin{center}
 \begin{figure}[htbp]
 \centering
 \includegraphics[width=0.85\textwidth]{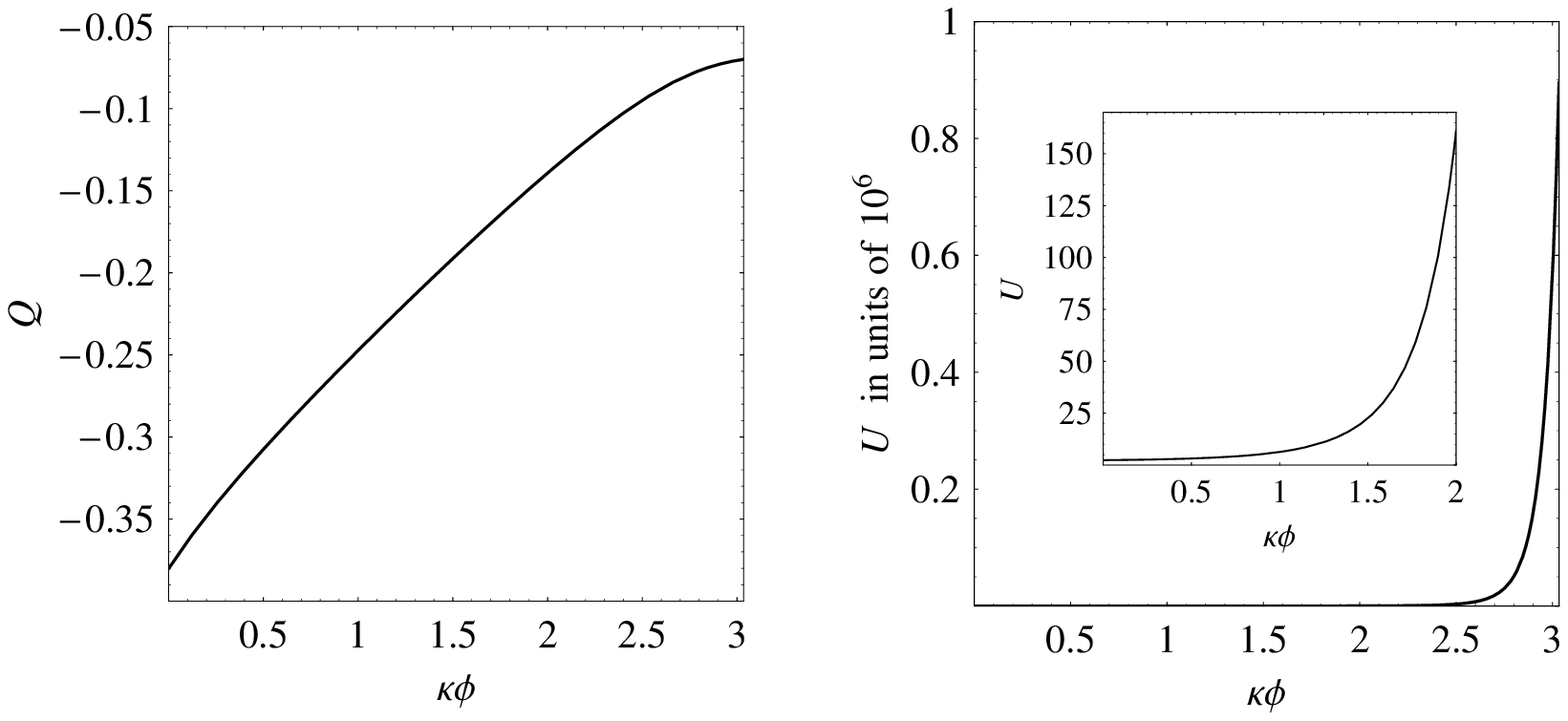}
 \caption{\label{fig4}
 $Q$ and $U$ as functions of $\kappa\phi$.
 See the text for details.}
 \end{figure}
 \end{center}


\vspace{-11mm} 


\section{Concluding remarks}\label{sec4}

In summary, we have shown that WDM, modified gravity and
 coupled CDM form a trinity, namely, they are
 indistinguishable by using the cosmological observations
 of both cosmic expansion history and growth history. In
 fact, they are three fairly different types of models: in the
 modified gravity models general relativity has been modified,
 while in both the WDM models and the coupled CDM models
 general relativity still holds. On the other hand, the
 interaction between CDM and dark energy is allowed in the
 coupled CDM models, while WDM and dark energy do not interact
 in the WDM models. However, we show that they are really
 indistinguishable by using the observations of both cosmic
 expansion history and growth history. This is not good
 news for the extensive attempts made in the literature
 to distinguish them (see e.g.~\cite{r3,r4,r5,r6,r30}). In
 particular, the indistinguishability of modified gravity
 and coupled CDM was shown in the previous works~\cite{r9,r10},
 while the indistinguishability of WDM and modified gravity,
 as well as the indistinguishability of coupled CDM and WDM,
 are shown in the present work. Therefore, to distinguish
 them, it is required to seek some complementary probes
 beyond the ones of cosmic expansion history and growth
 history (for instance, the observations on the
 small/galactic scale).

Some remarks are in order. Firstly, the models considered in
 this work are simple in fact. There are more complicated
 cases. For example, in the coupled CDM model, the role of
 dark energy can be played by other dynamical candidates
 rather than quintessence. On the other hand, the interaction
 form considered in the coupled CDM model is the type
 $\propto Q\rho_m\dot{\phi}$, which in fact can be more
 complicated in the literature. Secondly, the situation can
 be more complicated by extending these models. For instance,
 we can allow that there is also interaction between WDM
 and dark energy in the WDM model, or even allow CDM/WDM
 non-minimally coupled with gravity in all three of these
 types of models. Of course, one can also replace CDM with WDM
 in the modified gravity models. In these more complicated
 cases, it is more difficult to distinguish them by using the
 observations of cosmic expansion history and growth history.
 Thirdly, to distinguish these models, the other complementary
 probes beyond the ones of cosmic expansion history and growth
 history are desirable. For example, we can consider the
 observations on the small/galactic scale in which the
 structure formation is non-linear. The local tests of gravity
 on Earth or in the solar system are also useful, since general
 relativity can be tightly tested here. Of course, the high
 energy experiments are helpful too. For example, one might
 find the evidence of extra dimensions (required by some kinds
 of modified gravity models) in the very high energy colliders
 (e.g. the LHC at CERN). Finally, from the statistical point
 of view (e.g. Bayesian analysis), model selection is a rather
 subtle subject, in which priors and the number of parameters
 play an essential role (we thank the anonymous referee for
 pointing out this issue). The simplest model with the fewest
 free parameters and the minimal assumptions is favored.
 However, such a selection is based on the available data
 under consideration. When the other new data are added,
 the conclusion might be changed correspondingly. So, to
 distinguish the three types of cosmological models considered
 here, seeking the other complementary probes is still the
 best way. Recently, many significant progresses have been
 made in the observations on the small/galactic scale, the
 local tests of gravity, and the very high energy colliders
 (e.g. the LHC at CERN). It is hopeful to distinguish these
 models in the near future.


\section*{ACKNOWLEDGEMENTS}
We thank the anonymous referee for quite useful comments and
 suggestions, which helped us to improve this work. We are
 grateful to Professors Rong-Gen~Cai, Shuang~Nan~Zhang,
 Xiao-Jun~Bi for helpful discussions. We also thank Minzi~Feng,
 as well as Lixin~Xu, Long-Fei~Wang and Xiao-Jiao~Guo,
 for kind help and discussions. This work was supported in
 part by NSFC under Grants No.~11175016 and No.~10905005,
 as well as NCET under Grant No.~NCET-11-0790, and the
 Fundamental Research Fund of Beijing Institute of Technology.

\renewcommand{\baselinestretch}{1.1}


\end{document}